\documentclass [
  sigconf,
  nonacm=true
] {acmart}

\usepackage {
  nameref,
  cleveref,
  csvsimple-l3,
  enumitem,
  forest,
  makecell,
  setspace,
  listings,
  pgfplots,
  pgfplotstable,
  roboto-mono,
  tikz-uml,
  url,
}

\pgfplotsset {compat = 1.9} 
\usetikzlibrary {
  shapes.multipart,
  positioning,
  patterns,
  angles,
  er,
  arrows,
  fit,
  arrows.meta
}

\usepackage {custom-tikz}

\crefname {section} {Sect.}   {sections}
\Crefname {section} {Section} {Sections}
\crefname {figure}  {Fig.}    {figures}
\Crefname {figure}  {Figure}  {Figures}
\crefname {table}   {Tbl.}    {tables}
\Crefname {table}   {Table}   {Tables}

\newcommand \thdr [1] {\textbf {#1}}

\copyrightyear{2024}
\acmYear{2024}
\setcopyright{rightsretained}
\acmConference[MODELS Companion '24]{ACM/IEEE 27th International Conference on Model Driven Engineering Languages and Systems}{September 22--27, 2024}{Linz, Austria}
\acmBooktitle{ACM/IEEE 27th International Conference on Model Driven Engineering Languages and Systems (MODELS Companion '24), September 22--27, 2024, Linz, Austria}\acmDOI{10.1145/3652620.3687811}
\acmISBN{979-8-4007-0622-6/24/09}

\begin {document}

\title
  {From a Natural to a Formal Language with DSL Assistant}

\author{My M. Mosthaf}
\email {mymo@itu.dk}
\affiliation {
  \institution {IT University of Copenhagen}
  \city {Copenhagen}
  \country {Denmark}
}

\author {Andrzej Wąsowski}
\email {wasowski@itu.dk}
\affiliation {
  \institution {IT University of Copenhagen}
  \city {Copenhagen}
  \country {Denmark}
}

\begin {abstract}
  The development of domain-specific languages (DSLs) is a laborious and iterative process that seems to naturally lean to the use of generative artificial intelligence. We design and prototype \emph {DSL Assistant}, a tool that integrates generative language models to support the development of DSLs. DSL Assistant uses OpenAI's assistant API with GPT-4o to generate DSL grammars and example instances. To reflect real-world use, DSL Assistant supports several different interaction modes for evolving a DSL design, and includes automatic error repair. Our experiments show that DSL Assistant helps users to create and modify DSLs. However the quality of the generated DSLs depends on the specific domain and the followed interaction patterns.
  \looseness -1
\end {abstract}


\maketitle

\section {Introduction}

Modeling the world is a common concern in software development and the driving force in model-driven engineering (MDE). An MDE application is designed to address a subject area, a domain, and experts within this area.  Programs (models) are specified in domain-specific languages (DSLs) and opposed to general-purpose languages (GPLs), these have a higher level of abstraction, due to reduced expressiveness and syntax\,\cite{fowler.parsons:2011,dsl.design}. The syntax is familiar to domain experts, who are not necessarily developers.
\looseness -1

Meanwhile, the field of AI, especially generative AI and even more so the Large Language Models (LLMs), has since the beginning of this decade, revolutionized the world of software by narrowing the gap between complex (human) input and the creation of quality output. By speaking the language of the involved stakeholders, generative AI tools, e.g.\ Github Copilot\,\cite{Peng2023}, have been shown to increase productivity of software developers.

Generative AI may have a positive impact on MDE, especially when bridging between the domain experts and MDE tooling.   Many developers find translating the knowledge of domain experts into a domain-specific language definition to be a daunting task.  This perception is based primarily on many years of experience of teaching this to a broad range of university students.  At the same time, the expert language designers feel that the process followed is rather strict and streamlined, even repetitive.  This suggests that this predictable process could be supported with language-model-based tool.  In this  paper, we report on the design and prototype implementation of \emph {DSL Assistant}, a tool that integrates generative language models to support the development of external DSLs.  Specifically, DSL Assistant uses OpenAI's assistant API with GPT-4o to generate DSL grammars and example instances. To reflect real-world use, DSL Assistant supports several different interaction modes for evolving a DSL design, and includes automatic error solving.  In particular, we contribute:
\begin{itemize}

  \item A use case analysis, a design, and an implementation of DSL Assistant, a tool that can facilitate the development of DSLs and examples supported by LLMs.

  \item An extensive evaluation of the usability of the tool, and the capabilities of GPT-4o to solve tasks in this space.

\end{itemize}
DSL Assistant successfully helps users to create and modify DSLs. The quality of the generated DSLs depends on the domain and the interaction patterns.  GPT-4o is also remarkably effective in correcting its own mistakes when pointed out to it.
\looseness -1

\subsubsection* {Related Work}

Broadly in MDE, LLMs have been used to support model management\,\cite{Tang2019,Barriga2022,Dehghani2022,Weyssow2022}, code generation\,\cite{Nejjar2024,Xu:2024}, automating modeling tasks\,\cite{Chaaben2023,camara2023}, and generating recommendations\,\cite{Chaaben2023}. Cámara et al.\ analyze how ChatGPT performs at general modeling tasks across a number of domains and notations\,\cite{camara2023}. They conclude that the performance at the time was ``\emph{limited, with various syntactic and semantic deficiencies, lack of consistency in responses and scalability issues}.'' Kulkarni and coauthors present a symbiotic approach between MDE and generative AI similar to ours, but used for digital twin development not DSLs\,\cite {Barat2023}.  Very recently, Netz and coauthors investigated how LLMs can narrow the gap between domain experts and developers in the context of Web Application development\,\cite{michael2024}.  They use the MontiCore framework, while we use Xtext (for which much more LLM training material is available).  Busch et al.\ \cite{busch2024} present an approach to no-code \emph {graphical} DSL application development exploiting ChatGPT to generate code. We are interested in building an LLM-based textual language workbench.  Wang et al.\ \cite{cao2023} propose grammar prompting: an approach that enables LLMs to perform better in DSL generation tasks with in-context learning, expressing domain-specific constraints with grammars in Backus-Naur Form. Their experiments cover only a part of our use cases (cf.\ RQ3's last part in \cref{sec:evaluation}).
\looseness -1

\section {Background}\label {sec:background}

\subsubsection* {Languages, Models, and Meta-models.}

An origamist can write a paper crane tutorial and a paper hat tutorial in an origami tutorial DSL.\@ The underlying models of the tutorial examples conform to the grammar (syntax) of the origami DSL.  A DSL has both \emph {syntax} and \emph {semantics} concerning its structure and meaning, respectively. The syntax consists of rules that check whether a DSL's examples (instances) are structurally valid. For example, an origami tutorial must have a name and a number of steps, but an author is optional. These rules can be represented abstractly and concretely. The \textit{abstract syntax} is represented in Abstract Syntax Trees (AST) and is often specified as a meta-model (say Ecore) stored in computer memory. The \emph {concrete syntax} is visible for the user and depends on whether the DSL is graphical or textual, or both. A concrete syntax for a textual DSL is typically specified with regular expressions or context-free grammar (like Xtext).  We mostly work with concrete syntax below.
\looseness -1

\subsubsection* {Large Language Models}

Language Models (LLMs) are machine learning models efficient in a number of text-oriented tasks  such as customer service, idea generation, proofreading, and especially code generation. LLM technologies are revolutionizing  the (commercial) market today, but not without challenges. OpenAI has trained a wide range of models with different capabilities, speeds, and prices per token.\footnote{https://platform.openai.com/docs/models} The models convert input tokens to output tokens in a likely but non-deterministic way. For text-to-text generation, OpenAI offers four main models: GPT-3-turbo, GPT-4, GPT-4o (May 2024), and GPT-4o mini (July 2024). Quality-wise, GPT-4 and GPT-4o are competitors, but GPT-4o is faster and better than GPT-3-turbo.  We have not experimented with GPT-4o mini as it was released after this paper has been written.
\looseness -1

Text generation is via OpenAI's \emph {chat completion} and \emph {assistant} APIs.\footnote{https://platform.openai.com/docs/guides/text-generation} Using the chat completion, one can send a message (a prompt) along with an (optional) model context (a list of messages) to a specific model and get a message (an answer) back.  It is not possible to continue the conversation in subsequent API calls, unless the entire previous conversation is used as context. Assistants, on the other hand, resemble the functionality of ChatGPT, in the sense that they can access persistent threads. An assistant has a model and a number of settings, including predefined instructions to tune its answer. Messages (prompts and answers) are sent through threads which hold the context. After sending prompts to a thread of an assistant, one has to run the thread and wait for completion to retrieve the answer.  Like with chat completion, it is also possible to provide initial messages to a thread, i.e., providing additional model context. Both APIs can produce structured output in JSON format.
\looseness -1

\section {Analysis and Requirements}%
\label {sec:analysis}

DSL Assistant aims to help users develop DSLs more easily by exploiting capabilities of LLMs. In this work, we focus solely on concrete syntax development. We consider the following use cases.
\looseness -1

\subsubsection* {Origami Tutorial}

Alice has previously made illustrated step-by-step origami tutorials, following a manual, time-consuming, and error-prone process.  She wants to support her workflow with a DSL-based tool.  She has identified the purpose, stakeholders, concepts, relations, and examples of the Origami Tutorial domain, and she can describe these concisely. She has limited programming experience, and creating a DSL poses a significant challenge for her.  Alice's simplified process could be as follows.
\begin {enumerate}

    \item Alice prompts the language model with an example: a tutorial title, the size and shape of the initial paper sheet, and some folding steps. The assistant proposes the \emph{root\,version} of a DSL.\@
    \looseness -1

    \item She extends the DSL by telling the LLM how to obtain points and lines with geometrical operations within each step, and how to fold a paper with a mountain fold using a set of valid combinations of points and lines to find the location of the fold, and which new points and lines a fold can result in.
    \looseness -1

    \item Alice is not content with the new \emph {version}, so she deletes it and backtracks to the root version, explaining the concept more precisely this time.

    \item Alice realizes that she forgot to mention valley folds and extends the latest DSL version manually; while doing this, she introduces a syntax error in the grammar definition.

    \item DSL Assistant attempts to correct the error, and it succeeds after two attempts, producing a corrected version of the DSL.\@

    \item Prompted by Alice, the DSL Assistant creates a new example of a tutorial on how to make a paper frog, complying with the latest version of the DSL.\@

    \item Alice does not like the math-like syntax (such as equality symbols) and changes the DSL again, asking the assistant to replace these symbols with more common symbols, and words like `grab' and `fold.'

    \item Alice manually creates a new example of a paper hat tutorial for the last version of the DSL.\@ \qed

\end {enumerate}

\subsubsection* {Inventory Management}

Bob is responsible for streamlining his company's room booking infrastructure. He has decided to define a DSL for each task.  For the first DSL, Bob has collected a lot of real examples of meeting room booking from the employees, for instance: \emph {Birgitte, Benjamin, and I want to book room A for a physical meeting tomorrow between 14 and 15.15 with remote participants.}  Bob's simplified process could look as follows.
\begin{enumerate}

    \item Bob translates some of the natural language examples to a DSL using the DSL Assistant's LLM.\@

    \item Independently, Bob creates examples manually by translating English to a structured language, for instance, \emph {Type: Physically/remote, When: 07/06/2024 at 14:00--15:00, Where: Room A, Who: Birgitte, Benjamin, and Bjørn}.

    \item Bob asks colleagues to select three best examples and prompts DSL Assistant to define the syntax based on them.

    \item Bob creates more examples by asking DSL Assistant to translate some of the collected ones to the newly created syntax.
    \looseness -1

\end{enumerate}
For the second DSL, Bob has categorized the inventory items as IT, furniture, and other. He created a new DSL manually, where each item is defined with a name, a description, and a category with newlines in between. Every property name and value are separated by a colon. After some time, the company discovered that the inventory category is sometimes mistyped, and therefore they want to predefine the categories. Since Bob is no longer there, Benedikte, has to update the DSL.\@
\begin{enumerate}[resume]

    \item Benedikte opens the tool and then the inventory DSL project.

    \item Benedikte extends the DSL by prompting DSL Assistant to add predefined inventory categories `IT',  `Furniture,' and `Other' instead of a general character string. \qed

\end{enumerate}

\noindent
The \emph {target audience} of the tool are any stakeholders in DSL projects, which includes software engineers (e.g., Bob), domain experts (e.g., Alice), and business experts (e.g., Benedikte). The level of software development expertise can vary from novice to expert.  Select features may be targeted against a certain user type, for instance, manual code editing for users with programming skills.

Users can navigate the tool and manipulate data and meta-data. A user can create and access a project. Within a project, a user can view, delete, create, and update (improve, extend, modify, change, correct, etc.) both DSLs and examples. DSL Assistant uses a LLM to assist the user in creating and updating tasks.  A DSL and an example each have a definition, which depends on their concrete implementation. A definition can either be correct or faulty.  An error message explaining the problem is provided for faulty definitions, so that it can be corrected, either manually or automatically.

The development process is typically non-linear, therefore, the tool tracks \emph {versions} of the DSLs and the examples, that can facilitate a flexible process. When a user updates a DSL or an example, a new version of the DSL/example is instantiated and traced to the ancestor, a \emph{base}, to facilitate backtracking.  Similarly, when a user creates a new DSL syntax (for an example) or an example (for a DSL) the two involved definitions are traced. The co-evolution of models---in our case the DSLs and examples---has great potential in improving both quality of the models and stakeholder communication\,\cite{edm}.
\looseness -1

\section {The Implementation of DSL Assistant}%
\label {sec:implementation}

\input {fig/spread.tex}

\subsubsection* {Architecture}

DSL Assistant is implemented as a Web Application based on an Open AI LLM and a headless installation of the Xtext language workbench. Its architecture consists of three layers (presentation, logic, and data) and the communication between these (\cref {fig:architecture}). The primary entity of each layer accesses shared Typescript types and functions, to ensure consistency across layers.

The \emph {presentation layer} is a web front-end written in HTML, CSS, and Typescript using the web component library Lit (\cref {fig:gui}). The front-end communicates directly with the database when querying initial entity data and deleting entities. Otherwise, it sends a request with user input to the application server. The \emph {logic layer} uses two web servers---an application server and an Eclipse server that both expose a REST API.\@ The application server is a standard Node.js server using the Express library. The Eclipse server administrates an Eclipse Xtext project. The  application server creates new entities in response to requests. For some requests, it reads from the database and communicates with the language model. The entities created in this process are inserted into the database, validated with the Eclipse server and possibly returned to the web front-end (e.g.\ as a new version). See also \cref {fig:version_processing}.  The \emph {data layer} uses a PostgreSQL database through the Supabase platform.

\subsubsection* {Definitions and Versions}

The definition of an entity---a DSL or an example---is represented as a \textit{concrete-syntax}. DSL versions are defined by grammars in Xtext format and examples are defined by the sentences derived from the grammar. An abstract syntax is derived from concrete syntax and presented in Ecore to visualize the DSL designs.  It is not an interactive editable object at this point.

\looseness -1
DSL Assistant tracks versions of definitions behind-the-scenes. The exact messages (prompts to and answers from the LLM) are hidden from the user. DSL Assistant manages the conversation sessions linked to versions.  A version has three key attributes (\emph {kind}, \emph {input format}, and \emph {base}) and an \emph {input} (the prompt). Two \emph {kinds} are supported: make a DSL or an example. The \emph {input format} defines what format of data that the version's definition should be constructed from: a formal \emph {definition} (grammar, example), a natural language description (\emph {properties}), or an \emph {error message} from the language workbench (used to correct faulty versions).  Each new version is linked to a \emph {base} version (unless it starts a new development, so it is a root version).  The possible prompt configurations are summarized in the feature model of \cref{fig:feature_model} along with the following constraints.
\begin{enumerate}

    \item A version may have more than one base iff it generalizes several existing versions. It must be then a DSL syntax version and the base contexts must be examples.
    \looseness -1

    \item An error message input must be dependent on a faulty base.

    \item For an error message input the kind must be the same both for the present version and for the base.

    \item A new version  must have the same kind as the base and the base must be the latest version known, that is, the base must have no other successors. (Without this constraint, the tool would have to ensure version naming/numbering for referrals).

\end{enumerate}
Under these constraints the number of valid configurations of the model is twelve. DSL Assistant translates each valid option combination into a prompt and sends it to the LLM, which processes it and responds with an answer. The answer is translated into a version linked into the version graph. Given the non-deterministic nature of the LLMs, the answers are not guaranteed to be correct.
\looseness -1

\subsubsection* {Large Language Model and Prompting}

We use OpenAI's GPT-4o, and a single assistant, not a chat completion, as the assistant's API threads are useful to preserve context. The assistant works in the JSON mode and uses the following set of instructions.
\begin{enumerate}

    \item Omit the human-oriented text such as intro, summary, additional properties: \emph {Don't justify your answers. Don't give information not mentioned in the CONTEXT INFORMATION}.

    \item Build a JSON output \emph {Return answer as JSON format, within the properties specified in the prompt}.

    \item Formatting of code: \emph {Always return code in plain text, that is, no markdown}.

    \item Vary answers when correcting errors: \emph {If there are mentioned errors in the result, carefully read through the errors and try to CHANGE the result and do not just return the same result}.
    \looseness -1

\end{enumerate}
To simplify the communication with GPT-4o, the conversation is always one prompt and a single answer. Regardless of the input, a version of a grammar is always constructed by GPT-4o. For instance, a grammar input will be processed and (hopefully) reproduced by the LLM.\@ In that way, each version is related to a certain conversation state (and vice versa).  Occasionally, we end up in use cases where there is no grammar definition in the base, or no base, but we work with a grammar anyways.  Then we include the definition of the grammar as a context in the initial prompt (for instance copied from another conversation).

The sessions are following GPT-4o assistant's threads. If there is no base or a base has no context, a prompt is introduced in a new thread. Otherwise, the existing thread of the base is continued with a new prompt.
After a run on the thread is completed, the last GPT-4o answer is retrieved. Thanks to the constraint four in the feature model, one thread always concerns versions with the same kind of entity, and thus, the latest version on that thread is the only one that can be a base with context for a new version. In that way, GPT-4o never needs to be told which version is referred to in a thread.
\looseness -1

A prompt consists of an introduction, context, input data, and output indicator\,\cite {Giray2023} (cf.\,\cref{fig:prompts}).  The introduction guides the behavior of GPT-4o towards a desired answer (\emph {Return a grammar (Xtext) for a DSL and a name and a description of this DSL}).  An input data defines the task matter (\emph {The grammar should encapsulate the following properties: <DSL properties>} or \emph {The grammar should be generalized from the following instance: <example text>}).  An output indicator contains formatting instructions (\emph {Output the grammar in a `grammar' property, and the name in a `name' property, the description in a `description' property}).  The context element is responsible for providing the base (\emph {This is a grammar (Xtext) for a DSL:\@ <DSL grammar>}), if such is not available in the base.
\looseness -1

\begin {figure*} [t]
\begin {minipage} [t] {.49 \linewidth}
\begin {lstlisting}
Return a grammar (Xtext) for a DSL and a name and a description of this DSL

The grammar should encapsulate the following properties:

A DSL for defining an origami tutorial. An origamist can define their tutorial with name, authors, paper and steps, where a step has a description and folds to be made. The different kinds of folds are defined in the Yoshizawa-Randlett system and they have to follow the rules defined with the Huzita-Hatori axioms.

Output the grammar in a 'grammar' property, and the name in a 'name' property, the description in a 'description' property
\end{lstlisting}
\end{minipage}
\hfill
\begin {minipage} [t] {0.48 \linewidth}
\begin{lstlisting}
Something went wrong, this is the error:

Error raised during compiling of Xtext grammar
XtextSyntaxDiagnostic: null:2 required (...)+ loop did not match anything at input '<EOF>'

Carefully read error and try to find and solve the mistake and return the new corrected result

Output the grammar in a 'grammar' property, and the name in a 'name' property, the description in a 'description' property

Return how the new result is corrected in an 'adjustment' property
\end{lstlisting}
\end {minipage}

\vspace {-2mm plus 0.5mm minus 1mm}

\caption{Example prompts used by DSL Assistant. Left: A prompt for a DSL kind, properties input, and no base.  Right: A prompt constructed for a DSL kind with an error message input and a faulty DSL version as a base}%
\label{fig:prompts}

\end {figure*}

\subsubsection* {Validation and Repair}

When the kind is a DSL, the produced concrete context (grammar) is forwarded to the Eclipse server to test a build. A produced Ecore file  is extracted from the build, depending on the outcome, and a new version is instantiated and inserted into the database. If the build failed and an error message is available, the message is forwarded to GPT-4o to attempt a grammar repair (up to four times). See the right side of \cref{fig:prompts} for an example prompt.

The problems with faulty grammars manifested either at compile-time (generation-time) or at run-time.  The compile-time problems included broken syntax (missing, mismatched, or invalid symbols), linking errors (non-resolvable references to rules, packages, and types), and transformation errors (unknown or wrongly used types, rules and packages). The runtime-time exceptions observed when the Java program was run after compilation included invalid state (duplicated rule, missing start parsing rule, invalid cross-references), missing files (unknown elements), and null pointer exceptions when using Ecore reflection API.\@
\looseness -1

\begin{table}[t]

  \begin {tabular} {
    >{\small}r
    >{\small}l
    >{\small}r
    >{\small}r
  }

    \thdr {\#}
    & \thdr {Name}
    & \thdr {\llap{Description}}
    & \thdr {GPTzero \%}
    \\
    \hline

    \begin{filecontents*}{data/domains.csv}
index,name,dslName,origin,aiPropability
01,Smart Home Automation,HomeScript,LLM-made,\textbf{47}
02,Supply Chain Optimization,SupplyChainDSL,LLM-made,92
03,Genetic Research,GeneScript,LLM-made,98
04,Legal Contracts,LegalDSL,LLM-made,100
05,Urban Planning,CityPlanDSL,LLM-made,100
06,Cybersecurity,SecScript,LLM-made,100
07,Agricultural Management,AgriScript,LLM-made,98
08,Mental Health Therapy,TherapyDSL,LLM-made,98
09,Renewable Energy Management,RenewableDSL,LLM-made,100
10,Education and Learning Analytics,EduScript,LLM-made,100
11,Fashion Design and Manufacturing,FashionDSL,LLM-made,100
12,Environmental Monitoring,EnviroDSL,LLM-made,100
13,Origami Tutorial,OrigamiTutorialDSL,Man-made,3
14,Knitting pattern,KnittingPatternDSL,Man-made,2
15,Crossword,CrosswordDSL,Man-made,23
16,House Plant Management,HousePlantManagementDSL,Man-made,3
17,Recipe,RecipeDSL,Man-made,\textbf{44}
18,LEGO Assembly Kit,LegoKitDSL,Man-made,\textbf{44}
\end{filecontents*}
\csvreader [
      head to column names = true,
      head to column names prefix = T,
    ] {data/domains.csv} {}
    {\Tindex & \Tname & \Torigin & \TaiPropability \\}
  \end{tabular}

  \vspace {-2mm}

  \begin {minipage} {\linewidth}
  \footnotesize \textbf {Smart Home Automation:} A DSL designed to create, customize, and automate smart home workflows. It would allow users to define rules and scenarios for devices (lights, thermostats, security cameras, etc.) to interact, such as 'if the temperature drops below 68°F, turn on the heating' or 'turn off all lights and lock doors when I say Goodnight'.
  \end {minipage}

  \medskip

  \caption{Subject domains for grammar generation, along with one example description}%
  \label{tab:domains}

\end{table}


\section {Evaluation Design}%
\label {sec:evaluation}

We ask the following research questions and answer them in three experiments, one involving human subjects and two automatic.
\begin {itemize}

  \item [RQ1] How do users perform using DSL Assistant? How does DSL Assistant facilitate this interaction?

  \item [RQ2] Is GPT-4o able to generate correct Xtext grammars? How effective it is at correcting Xtext grammars?

  \item [RQ3] Is GPT-4o able to instantiate examples from Xtext grammars? Can it generalize Xtext grammars from examples?

\end {itemize}

\subsubsection* {RQ1}\label {sec:met-a}

We observe four subjects interacting with DSL Assistant (one at a time, without inter-subject influence) using a think-aloud protocol. Each solves 17 tasks on a DSL for the game of chess while interacting with DSL Assistant in English.  The tasks are mostly derived from configurations in \cref{fig:feature_model}, which define what kind of input  and artifact is being manipulated.  The complete list of tasks is included in the thesis of Mosthaf\,\cite {mosthaf:2024}.  A subject points out to a researcher their actions and observations.  They are allowed to ask for clarification or hints.  The questions, hints, observations, and, unexpected behaviors of DSL Assistant are noted down. The protocol is documented with a screen and audio recording. Each subject elaborates on their experience in a follow-up interview. We ask questions within three categories: semantics, features/quality attributes, and overall experience\,\cite {mosthaf:2024}. The interview is recorded. The protocol has been pilot-tested on a CS master student with shallow modeling experience (no prior modeling course). The collected data is coded, with focus on the behavior of both the users and DSL Assistant.
\looseness -1

All four subjects are Danish students enrolled in an English Computer Science Master's at the IT University of Copenhagen and have passed the 7.5 specialization course on \emph {Modelling Systems and Languages}, largely following the text book by Wąsowski and Berger\,\cite {dsl.design}. One had taken the course in 2023, the others passed the exam a couple of days before the experiment. The course uses the same technologies (Xtext and Ecore) as DSL Assistant.
\looseness -1

\subsubsection*{RQ2}\label{sec:met-c}

We use DSL Assistant to automatically create grammars in one shot. Then we check whether these grammars are syntactically (using Xtext) and semantically correct (manually). The faulty  grammars are fed into DSL Assistant's automatic repair process in two modes: with the prior conversation context and without the context---in both cases the broken grammar and the error message are provided.  At most 5 attempts to repair each grammar are made.

We use 18 test domains: twelve with LLM-made and six with man-made descriptions (\cref{tab:domains}).  An example domain description is shown in the bottom of the table; the complete  list is included by Mosthaf\,\cite {mosthaf:2024}. GPT-3.5 has been used to select and generate the LLM-made domain descriptions. The first author invented the man-made ones.  We used GPTZero, which calculates the probability of a text to be LLM-generated\,\cite{Walters:2023}, in order to see whether there is a difference between these two populations.  The scores show that the domain descriptions are indeed different from a language model perspective.
\looseness -1

\begin {table} [t]

  \newcommand \mcs {\hspace {3pt}}

  \begin {tabular} {
    @{}
    >{\small}r
    @{\mcs}
    >{\small}l
    @{\mcs}
    >{\small}l
    @{\mcs}
    >{\small\raggedright\arraybackslash}p{47.7mm}
    @{}
  }

      \textbf{\#}
    & \thdr {Domain}
    & \thdr {Source}
    & \thdr {Examples}
    \\
    \hline

    \begin{filecontents*}{data/dsls.csv}
index,origin,domain,examples
1,\cite{dsl.design}, Fsm,               {\emph {coffee maker}}; complete login; simple login
2,\cite{dsl.design}, Robot,             {\emph {random walk}}; no events
3,course\,exam,      Origami\,tutorial, {\emph {paper hat tutorial}}; crane tutorial
4,course\,exam,      Store,             {\emph {danish supermarket}}
5,course\,exam,      School,            {\emph {small\,middle\,school\,1}}; small\,middle\,school\,2
6,course\,exam,      School query,      {\emph {four day workweek}}; missing teacher; pre- pa\-ra\-tion; too many hours; free thursday; not enough hours;\,not\,right\,hours
\end{filecontents*}
\csvreader [
      head to column names = true,
      head to column names prefix = T,
    ] {data/dsls.csv} {}
    { \Tindex
    & \textls[-10]{\Tdomain}
    & \textls[-10]{\Torigin}
    & \textls[-10]{\Texamples} \\}

  \end{tabular}

  \vspace {-2mm plus 0.5mm minus 1mm}

  \caption{DSLs and examples used for example generation test}%
  \label{tab:dsls}

\end{table}


\subsubsection* {RQ3}\label {sec:met-e}

To answer the first sub-question we create an example version for a known grammar and matching verbal example descriptions of a known formal instance (ground truth). We introduce an Xtext grammar into the system from an original source (no LLM-generation) as an initial version without a base. Then we attempt to synthesize an example, following three kinds of example descriptions: a \emph {general} description of one sentence of ten or less words, a \emph {non-technical} description---a longer human-targeted explanation that does not mention grammar rules, and a \emph {technical} description incorporating grammar rules and the exact data to consider.  We manually analyze whether the example synthesis was successful.  Notice that we do not judge whether the instances are better or worse than the proposed ground truth.  We merely check whether each description managed to allow the tool to reconstruct the ground truth instance.  We also recognize the risk of bias in us writing the proposed descriptions.

To answer the second sub-question we generalize two grammars for a DSL:\@  the first based on four similar examples as a base (the ground truth and the three synthesized one in the previous paragraph), and the second based on all available examples for a DSL as a base.  If the synthesized grammar turns out to be faulty, up to four attempts are made to repair it, using the same procedure as in RQ2.  We assess the outcome by comparing the obtained grammar manually to the grammar in the original source.

We conduct the experiment using six DSLs (\cref{tab:dsls}), a number of examples (one primary) for each DSL, and three descriptions (a general, a non-technical, and a technical description) of the primary example. The descriptions of the primary example have been formulated in English by the first author.  All examples are based on publicly available materials, in principle accessible for training of GPT models.  The collection includes both structural and behavioral languages. Details are available in the thesis report\,\cite{mosthaf:2024}.

\section {Results}\label {sec:results}

\subsubsection* {RQ1}\label {sec:res-a}

All test subjects were able to navigate and handle data rather trouble-free. Throughout the session, the subjects showed signs of maturation and explicitly stated how they became better at using DSL Assistant; faster and more familiar with the concepts. DSL Assistant was able to generate content automatically, facilitate an iterative non-linear process, and eliminate cumbersome use of the Eclipse IDE.\@ The subjects agreed that the automatic repair worked well. Two subjects remarked that DSL Assistant would have been a great aid during the modeling course.

The usability problems concentrated around the location of UI elements and long response times. The concepts of version, base, context, and properties were cognitively confusing, and it could be beneficial to further hide them from the users. Some subjects were confused whether the grammar or the meta-model were derived using a language model, and missed that only grammars are refined during the interaction, while meta-models are derived automatically. Subjects have been challenged to understand  exactly, how the new content was supposed to differentiate depending on the context options, which is consistent with the confusion about the basic concepts. Admittedly, the large variability of interactions have been included in the DSL Assistant to facilitate broader experimentation with the LLM, so the subject confusion is justified. Subjects reported \emph missing features such as missing tool-tips, a better overview, sorting and filtering of versions, syntax highlighting in the grammar editor, and always showing an example for the current grammar.

\begin{figure}[t]

\pgfplotstableread[col sep=comma] {data/chart_ai.csv}\aidatatable

\pgfplotstableread[col sep=comma] {data/chart_human.csv}\humandatatable

    \pgfplotsset{every axis/.style={
        tickwidth         = 0pt,
        axis line style={draw=none},
        legend style={at={(0.5,-0.00)},
        legend style={draw=none},
        anchor=north,legend columns=-1},
        bar width=0.35cm,
        xbar stacked,
        width=1.0\linewidth-2.0cm,
        xmin=0,
        xtick style={draw=none},
        axis x line=none,
        enlarge x limits={upper, value=0.0},
        y dir=reverse,
        y=0.4cm,
        enlarge y limits={true, abs value=0.75},
        ytick=data,
        y tick label style={align=right,text width=3.1cm, font=\scriptsize},
        nodes near coords,
        nodes near coords align={horizontal},
        nodes near coords style={black},
        visualization depends on={y \as \rawy},
        every node near coord/.append style={
            anchor=east, 
            xshift=\pgfkeysvalueof{/pgf/bar width}/2, 
        }
    }}
    \centering
    \begin{tikzpicture} []
    \begin{axis}[
        yticklabels from table={\aidatatable}{domain}
    ]
    \addplot+ [
        xbar
        , pattern=north east lines
        , pattern color=gray
        , draw=black
    ]
    table[y=index, x=ok] from \aidatatable;
    \addplot+ [
        xbar
        , pattern=crosshatch
        , pattern color=gray
        , draw=black
    ]
    table[y=index, x=faulty] from \aidatatable;
    \end{axis}
    \node[above, font=\small, xshift=1.8cm] at (current bounding box.north) {LLM-made Definitions};

    \end{tikzpicture}
    \begin{tikzpicture} []
    \begin{axis}[
        yticklabels from table={\humandatatable}{domain}
    ]
    \addplot+ [
        xbar
        , pattern=north east lines
        , pattern color=gray
        , draw=black
    ]
    table[y=index, x=ok] from \humandatatable;
    \addplot+ [
        xbar
        , pattern=crosshatch
        , pattern color=gray
        , draw=black
    ]
    table[y=index, x=faulty] from \humandatatable;
    \legend{OK, Faulty}
    \end{axis}

    \node[above, font=\small, xshift=1.8cm] at (current bounding box.north) {Man-made definitions};
    \end{tikzpicture}

    \vspace {-2mm plus 0.5mm minus 1mm}

    \caption{1-shot grammar generation across domains}%
    \Description{Bar diagram}%
    \label{fig:auto-01-chart}

\end{figure}
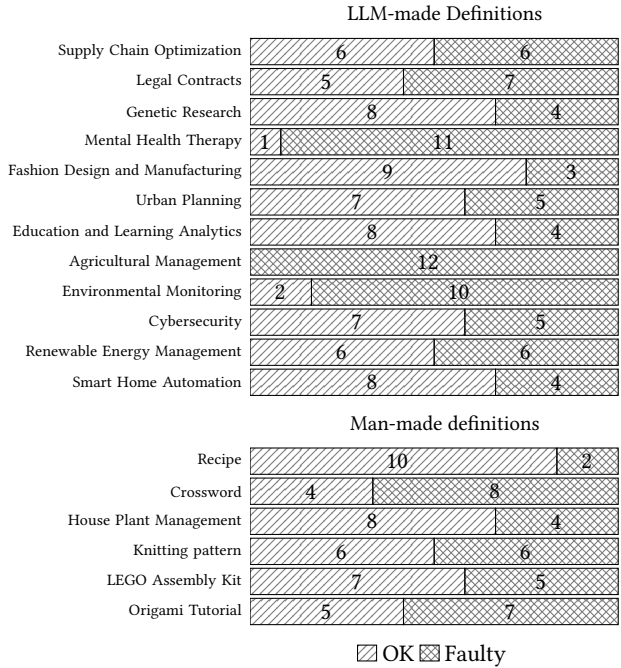

\begin {table} [t]

\newcommand \mcs {\hspace {3pt}}

\begin {tabular} {
  >{\small}l
  >{\small}r
  @{\mcs}
  >{\small}r
  @{\mcs}
  >{\small}r
           |
  >{\small}r
  @{\mcs}
  >{\small}r
  @{\mcs}
  >{\small}r
}

  & \multicolumn3{c}{\thdr{\small with    context}}
  & \multicolumn3{c}{\thdr{\small without context}}
  \\

    \thdr {Domain def.}
  & \thdr {\#succ}
  & \thdr {\%succ}
  & \thdr {\#attempts}
  & \thdr {\#succ}
  & \thdr {\%succ}
  & \thdr {\#attempts}
  \\

  \rlap {LLM-made}
  & 56
  & 77
  & 1.8 $\pm$ 1.06
  & 50
  & 68
  & 2.0 $\pm$ 1.02
  \\

  \rlap {man-made}
  & 22
  & 92
  & 1.3 $\pm$ 0.77
  & 21
  & 88
  & 1.4 $\pm$ 0.93
  \\

  \rlap {\thdr{overall}}
  & 78
  & \textbf {80}
  & 1.6 $\pm$ 1.01
  & 71
  & 73
  & 1.8 $\pm$ 1.02

\end{tabular}

\medskip

\caption{Automatic repair rates for faulty DSL grammars}%
\label{tab:auto-02-part2-agg}

\end{table}

\begin {table} [t!]

\begin {tabular} {
  >{\small}l
  >{\small}l
  >{\small}r
}

  \thdr{Exception}
& \thdr{Phase}
& \thdr{\#}
\\
\midrule

\begin{filecontents*}{data/error-2.csv}
category,error,count,fixed,percentage
Compile-time,\texttt{\scriptsize XtextSyntaxDiagnostic},224,75,30
Compile-time,\texttt{\scriptsize XtextLinkingDiagnostic},283,101,31
Compile-time,\texttt{\scriptsize TransformationDiagnostic},96,58,48
Compile-time,Total,\textbf{405},159,\textbf{35}
Run-time,\texttt{\scriptsize FileNotFoundException},8,1,17
Run-time,\texttt{\scriptsize IllegalStatementException},7,1,17
Run-time,\texttt{\scriptsize NullPointerException},1,0,0
Run-time,Total,\textbf{16},2,\textbf{14}
\end{filecontents*}
\csvreader [
head to column names = true,
head to column names prefix = T
] {data/error-2.csv} {}%
{\Terror & \Tcategory & \Tcount \\}

\end{tabular}

\vspace {-2mm plus .5mm minus 1mm}

\caption{Faulty grammars by error kind, as reported by Xtext}%
\label{tab:auto-02-error-2}

\end{table}

\subsubsection*{RQ2}
\label{sec:res-c}

GPT-4o successfully generated $216 = 18 \times 12$ grammars, half of these correct at the first shot (we asked for twelve results for each initial prompt). Interestingly, the tool succeeded a bit more often  with man-made descriptions of domains (56\%), than with the LLM-made (47\%). DSL Assistant failed to obtain a correct grammar for all descriptions in the Agricultural Managament domain (0\% success). \Cref{fig:auto-01-chart} gathers the one-shot success rates across the domains.
\looseness -1

GPT-4o successfully extracts the key aspects and relations in all domains and turns these into grammars, which can be transformed into corresponding meta-models with classes, attributes, and references. The grammars use enumerators, inheritance, and self-references. The productions describing structural aspects such as a list of materials, a title, a type, etc., have similar syntax in almost all the generated grammars. A line of such rule describes either a single property or a list of properties, for instance:
\begin{lstlisting}
'Pattern': title=STRING
'Steps': '{' steps+=Step (',' steps+=Step)* '}';
\end{lstlisting}
Colons and curly brackets are GPT-4o's preferred choices for syntax, even though it can be argued that these are not very domain-specific. On the other hand, the grammar rules describing behavioral aspects of a DSL, such as LEGO brick stacking, paper folding, etc., use simple syntax, often too simple to capture the behavior in a domain.
\looseness -1

Remarkably, DSL Assistant was able to automatically correct 83 of the faulty grammars---a success rate of 86\%, see \cref{tab:auto-02-part2-agg}. The repair rate was higher for man-described grammars, which bodes well for users; albeit this could also be a sign of a human bias in domain selection. When the session was continued (with context) instead of starting from scratch (without context) GPT-4o performed noticeably better.  However, as few DSLs were corrected only in the no-context mode, the combination of the two strategies seems to be the most effective (this is how the rate 86\% arises). In all the experiments, ChatGPT produced 421 distinct faulty intermediate grammars (most of them in the automatic repair process). \Cref{tab:auto-02-error-2} summarizes the different categories of errors observed.  In the table, we can see that by far most errors are statically detected by Xtext, which facilitates the repair process well.

To conclude, GPT-4o can generates syntactically correct DSL grammars in  a single shot about half of the time. The performance depends on the choice of a specific domain.  GPT-4o is able to capture key domain aspects in the grammars, but uses generic structure and symbols, while the behavioral relations are often too simple to be useful. Moreover, GPT-4o is capable of correcting the syntactically incorrect grammars in most cases. Note that these conclusions are made for 1-shot generation, without receiving feedback from users.
\looseness -1

\begin {figure*} [t]

\small
\newcommand \chdr [1] {\textbf {#1}}

\newcommand \lcol {.548 \linewidth}
\newcommand \rcol {.44 \linewidth}
\newcommand \lskip {-4pt}

\begin {minipage} [t] {\lcol}
\chdr{Ground truth~\cite{dsl.design}}

\smallskip

\begin{lstlisting}[lineskip = \lskip]
machine CoffeeMachine [
  initial ^initial
  state ^initial [
    on input "coin" output "what drink do you want?" and go to selection
    on input "break" output "machine is broken" and go to broken
  ]
  state selection [
    on input "tea" output "serving tea" and go to brewingTea
    on input "coffee" output "serving coffee" and go to brewingCoffee
    on input "timeout" output "coin returned; insert coin" and go to ^initial
    on input "break" output "machine is broken!" and go to broken
  ]
  state brewingCoffee [
    on input "done" output "coffee served. Enjoy!" and go to ^initial
    on input "break" output "machine is broken!" and go to broken
  ]
  state brewingTea [
    on input "done" output "tea served. Enjoy!" and go to ^initial
    on input "break" output "machine is broken!" and go to broken
  ]
  state broken
]
\end{lstlisting}
\end {minipage}
\hfill
\begin {minipage} [t] {\rcol}

\chdr{General:} A Coffee Machine FSM that can make coffee and tea.

\medskip

\begin{lstlisting}[lineskip = \lskip]
machine CoffeeMachineFSM [
  state Idle [
    on "coin" go to Ready
  ]
  state Ready [
    on "buttonCoffee" go to BrewingCoffee
    on "buttonTea" go to BrewingTea
  ]
  state BrewingCoffee [
    on "done" output "coffee" and go to Idle
  ]
  state BrewingTea [
    on "done" output "tea" and go to Idle
  ]
  initial Idle
]
\end{lstlisting}
\end {minipage}

\smallskip

\begin {minipage} [t] {\lcol}

\chdr {Technical:} A FSM called CoffeeMachine FSM. CoffeeMachine has five states, \^initial, selection, brewingCoffee, brewingTea, and broken. CoffeeMachine's initial state is \^initial. The \^initial state has three transitions: 1) with input ''coin'' and output ''what drink do you want?'' and the target is the selection state, 2) with input ''break'' and output ''machine is broken'' and the target is the broken state\ldots

\medskip

\begin{lstlisting} [lineskip = \lskip]
machine CoffeeMachine[
  state ^initial[
    on input coin output "what drink do you want?" and go to selection
    on input break output "machine is broken" and go to broken
  ]
  initial ^initial
  state selection[
    on input tea output "serving tea" and go to brewingTea
    on input coffee output "serving coffee" and go to brewingCoffee
    on input timeout output "coin returned; insert coin" and go to ^initial
    on input break output "machine is broken!" and go to broken
  ]
  state brewingCoffee[
    on input done output "coffee served. Enjoy!" and go to ^initial
    on input break output "machine is broken!" and go to broken
  ]
  state brewingTea[
    on input done output "tea served. Enjoy!" and go to ^initial
    on input break output "machine is broken!" and go to broken
  ]
  state broken[]
]
\end{lstlisting}
\end{minipage}
\hfill
\begin {minipage} [t] {\rcol}

\chdr{Non-technical:} A coffee machine where one can buy tea and coffee with a coin. Initially, one can insert a coin. After inserting a coin, one can select an beverage, either tea or coffee, if this takes to long it returns the coin. After selection the beverage kind, it will either serve coffee or tea. Anytime during the interaction with the coffee machine, it is possible to break the machine, so it no longer works.
\looseness -1

\medskip

\begin{lstlisting}[lineskip = \lskip]
machine CoffeeMachine {
 state Idle {
  on 'insertCoin' go to InsertCoin
 }
 initial Idle
 state InsertCoin {
  on 'selectBeverage' go to SelectBeverage
  on 'timeout' go to Idle
 }
 state SelectBeverage {
  on 'serveCoffee' go to ServeCoffee
  on 'serveTea' go to ServeTea
 }
 state ServeCoffee {}
 state ServeTea {}
 state Broken {}
}
\end{lstlisting}
\end {minipage}

\vspace {-3ex plus 1mm minus 2mm}

\caption{The original Coffee Machine example \cite{dsl.design} and the synthesized instances for three kinds of descriptions}%
\label{tab:examples-fsm}

\end{figure*}

\subsubsection* {RQ3}\label{sec:res-e}

Regarding the first sub-question, the instantiated examples use keywords and capture data mostly correctly according to the grammar rules, regardless of the used description kind.  \Cref {tab:examples-fsm} shows the outcomes for the Coffee Machine DSL along with the corresponding example texts.  We note minor deviations from the ground truth, such as pairs of empty brackets,  braces instead of brackets, wrong capitalization (not shown)   and missing dashes when listing opening hours (not shown). The most \emph {technical} description resulted in an example very close to the ground truth.  In examples for other DSLs, we noticed some unnecessary `hallucinated' expressions.  Inevitably, the texts that were the result of the \emph {general} descriptions differed the most from the original ground truth examples. They lacked data and contained wrong logic.

Regarding the second sub-question, we found that GPT-4o is able to generalize grammars from examples with rules abstracting over the keywords and data of the texts. The grammars also display minor syntactical problems. Seven out of twelve grammars were faulty, and two of these were repaired automatically within the four tries. For instance, the generalized School DSL based on the original example texts uses curly brackets instead of square brackets to surround the list of teachers, classes, etc. We have seen the cases of both  under-fitting or over-fitting (or both) compared to the ground truth. An example of under-fitting is found in the School DSLs: Teachers, classes, etc.,  are generalized to strings rather than their own classes as in the ground truth. One of the generalized Robot DSLs introduced notation for constant values of speed, angle, etc., even though these were not mentioned in the input examples. Furthermore, when generalizing from examples containing programming constructs, the performance is difficult to predict, and the logic of these may need to be validated afterward. It should be noted again that these conclusions are made for one-shot attempts. Obviously, the output can be improved in further interaction with the user.
\looseness -1

\subsubsection* {Validity}

We note several weaknesses regarding the internal validity. Semantic correctness of grammars and examples was established manually by non-domain experts. An automatic conformance check and grammar validity check could have increased the robustness of the answers to RQ3. The specific wording, tone, structure, and content of a prompt have undoubtedly influenced the quality of the answer. A study has shown that code performance varies considerably across prompt strategies\,\cite{hou2024}. The difference of lengths between the prompts constructed with and without context is significant, and might have impacted on the correctness of the produced grammars.
\looseness -1

We note risks to the generalization of results (the external validity). The quality of LLM output is affected by the specific textual representation and format of the input. However, the chance of obtaining a quality output is generally high for widespread formats\,\cite{Baumann2024}. To investigate how the choice of Xtext has influenced the answer, one would have to to experiment with other representations. The performance of different LLMs varies and in it is generally higher the subject, format, etc., are well-known and popular. Our results are biased to the selected domains, DSLs, and examples, and they could be different for other choices. We did take care to obtain a diverse data basis, mixing both LLMs and human creativity.
\looseness -1

\section {Conclusion}

We have presented DSL assistant, a tool that aims to help users can effectively manage DSL projects, evolve  DSLs and examples in concrete syntax.  DSL Assistant is based on GPT-4o, which has shown an ability to capture overall aspects of a domain in a grammar. The generated grammars are syntactically correct half the time and GPT-4o can automatically correct most of its incorrect outputs.  It also shows the ability to instantiate example texts from DSL grammars and to generalize DSL grammars from example texts, however with somewhat less precision.

The DSL Assistant only explores how GPT-4o performs at handling a fraction of the concepts known from MDSE. It could be interesting to extend the tool to support more MDSE concepts, such as abstract syntax, static semantics (e.g. OCL), dynamic semantics, and model transformations, so one could investigate its performance here. Furthermore, we have not yet studied the size of the domain (in terms of the domain model) that can be handled, given the GPT4o's training data and context window.  Neither we have experimented with other LLMs.  We speculate that performance in other languages than English will be much lower, but this also needs to be evaluated.  This could be particularly useful for education purposes.



\subsubsection* {Acknowledgements.}
We thank Adrian Hoff for a sparring session about an early implementation of the tool.

\bibliographystyle {ACM-Reference-Format}
\bibliography {main}

\end {document}